\documentclass[superscriptaddress,notitlepage,nofootinbib,letterpaper,english]{revtex4}

\usepackage{amssymb}
\usepackage{amsmath}
\usepackage{epsfig}
\usepackage{graphicx}
\usepackage{tabularx}
\usepackage{latexsym}
\usepackage{verbatim}
\usepackage{color}
\usepackage{braket}
\usepackage{multirow}
\usepackage{dcolumn}
\usepackage{mathtools}
\usepackage{placeins}
\usepackage{epstopdf}
\usepackage{wrapfig}
\usepackage[utf8]{inputenc}
\usepackage{hyperref}
\usepackage{diagbox}
\usepackage[normalem]{ulem}
\usepackage[english]{babel}
\usepackage{slashed}

\def\eq{\begin{eqnarray}}
\def\en{\end{eqnarray}}
\def\ep{\epsilon}

\renewcommand\sout{\bgroup \color{red} \ULdepth=-.5ex \ULset}

\DeclarePairedDelimiter\abs{\lvert}{\rvert}%
\allowdisplaybreaks[1] % n=0,1,2,3,4
\newcommand{\bm}[1] {\mbox{\boldmath{$#1$}}}
\newcommand{\half}{ {\textstyle\frac{1}{2}} }

\newcommand{\be}{\begin{equation}}
\newcommand{\ee}{\end{equation}}
\newcommand{\ba}{\begin{eqnarray}}
\newcommand{\ea}{\end{eqnarray}}
\newcommand{\sub}[1]{	\begin{subequations}
			#1
		     	\end{subequations} }

\newcommand{\la}{\langle}
\newcommand{\ra}{\rangle}

\usepackage[dvipsnames]{xcolor}

\begin{document}
\preprint{}

%Title of paper
\title{Gravitational form factors of a spin one particle}

\author{Maxim V.~Polyakov}
	\affiliation{Petersburg Nuclear Physics Institute, 
		Gatchina, 188300, St.~Petersburg, Russia}
	\affiliation{Institut f\"ur Theoretische Physik II, 
		Ruhr-Universit\"at Bochum, D-44780 Bochum, Germany}

\author{Bao-Dong Sun}
\email{sunbd@ihep.ac.cn}
\affiliation{Institute of High Energy Physics, Chinese Academy of Sciences, Beijing 100049, People's Republic of China}
\affiliation{School of Physics, University of Chinese Academy of Sciences, Beijing 100049, People's Republic of China}
%\noaffiliation

\begin{abstract}
We define the form factors of the quark and gluon symmetric energy-momentum tensor (EMT).
The static EMT is related to  the spatial distributions of energy, spin, pressure and shear forces.
They are obtained in the form of a multipole expansion. The relations between gravitational form factors and the generalised parton distributions are given. 
\end{abstract}
%\pacs{11.40.-q,13.60.Fz,14.40.Be}
\date{\today}
\maketitle

\section{Introduction}
\label{sec:Introduction}

The gravitational form factors (GFFs) contain the information of the spatial distributions of energy, spin, pressure and shear forces inside the system \cite{Polyakov:2002yz}.  
The GFFs are defined through the matrix elements of the symmetric energy-momentum tensor (EMT). More details can be found in the recent papers~\cite{Polyakov:2018zvc,Lorce:2018egm}. For spin one particles, the GFFs, or EMT FFs, have been discussed in the literature~\cite{Holstein:2006ge,Abidin:2008ku,Taneja:2011sy}, but, to our best knowledge, EMT-nonconserving FFs are  either not disscused~\cite{Holstein:2006ge,Abidin:2008ku} or incomplete~\cite{Taneja:2011sy}. Thus we introduced a definition for individual quark and gluon EMT FFs for spin one particles in Sec.~\ref{sec:Defination_of_EMT_form_factors}. 
In Breit frame, we find that that matrix elements of EMT can be expressed in terms of the multipole expansion for energy density, pressure and shear forces distributions, see Sec. \ref{sec:The-static-EMT-and-stress-tensor}. 
By considering the Mellin moments of the vector generalised parton distributions (GPDs), the sum rules between the GPDs and EMT FFs are found in Sec. \ref{sec:Sum-rules-GPD-and-GFF}. \\

The EMT of QCD can be obtained by varying the action $S_{\rm grav}$ of QCD 
coupled to a weak classical torsionless gravitational background field with 
respect to the metric $g^{\mu\nu}(x)$ of this curved background field according to~\cite{Belitsky:2005qn,Polyakov:2018zvc}
\ba
\label{Eq:EMT-from-gravity}
	\hat{T}_{\mu\nu}(x)=\frac{2}{\sqrt{-g}}\,
	\frac{\delta S_{\rm grav}}{\delta g^{\mu\nu}(x)}
\ea
where $g$ denotes the determinant of the metric  
{(the signature of the metric we use is $+---$)}. This procedure 
%(see e.g.\ App.~E of \cite{Belitsky:2005qn} for a pedagogical description) 
yields a symmetric Belinfante-Rosenfeld EMT. The quark and gluon 
contributions to the total EMT operator are given by 
{
\sub{
\label{eq:EMT-QCD}
\ba
	\label{eq:EMT-QCD-q}
	T^{\mu\nu}_q &=& \frac{1}{4} \bigg[ \overline{\psi}_q\biggl(
	-i\overset{ \leftarrow}{\cal D}{ }^\mu\gamma^\nu
	-i\overset{ \leftarrow}{\cal D}{ }^\nu\gamma^\mu
	+i\overset{\rightarrow}{\cal D}{ }^\mu\gamma^\nu
	+i\overset{\rightarrow}{\cal D}{ }^\nu\gamma^\mu\biggr)\psi_q
	-g^{\mu\nu}\overline{\psi}_q\biggl(
	-\frac{i}{2}\,\overset{ \leftarrow}{\slashed{\cal D}}{ }
	+\frac{i}{2}\,\overset{\rightarrow}{\slashed{\cal D}}{ }
	{\,-\,m_q}\biggr)\psi_q  \bigg] \ ,
	\\
	\label{eq:EMT-QCD-g}
	T^{\mu\nu}_g &=& F^{a,\mu\eta}\,{F^{a,}}_{\eta}{ }^\nu+\frac14\,g^{\mu\nu}
	F^{a,\kappa\eta}\,{F^{a,}}_{\kappa\eta} \ .
\ea} }
Here $\overset{\rightarrow}{\cal D}_\mu=\overset{\rightarrow}\partial_\mu+ig\,t^aA_\mu^a$ 
and $\overset{\leftarrow}{\cal D}_\mu = \overset{\leftarrow}\partial_\mu-ig\,t^aA_\mu^a$ 
with arrows indicating which fields are differentiated, 
$F^a_{\mu\nu}=\partial_\mu A^a_\nu-\partial_\nu A^a_\mu-g\,f^{abc}A^b_\mu A^c_\nu$
and the SU(3) color group generators satisfy the algebra 
$[t^a,t^b]=i\,f^{abc}t^c$ and are normalized as 
${\rm tr}\,(t^at^b)=\frac12\,\delta^{ab}$.
The total EMT is conserved
\be\label{Eq:EMT-cons}
	\partial^\mu\hat T_{\mu\nu} = 0, \quad \quad
	\hat T_{\mu\nu} = \sum_q\hat T_{\mu\nu}^q+\hat T_{\mu\nu}^g \; .
\ee

\section{Definition of EMT form factors}
\label{sec:Defination_of_EMT_form_factors}

We use the covariant normalisation
$\la p^\prime, \sigma^\prime|\,p, \sigma\ra=2p^0\,(2\pi)^3\delta^{(3)}(\bm{p^\prime}-\bm{p})\delta_{\sigma\sigma^\prime}$
of one-particle states, and introduce the kinematic variables 
$P= \frac12(p^\prime + p)$, $\Delta = p^\prime-p$, $t=\Delta^2$.
The EMT form factors of a spin-1 particle in QCD we define as\footnote{We chose the naming of the form factors in line with
the naming used in Refs.~\cite{Polyakov:2002yz,Polyakov:2018zvc} for spin-0 and spin-1/2 particles.}
\eq  \label{Eq:EMT-FFs-spin-1}
\langle p^\prime,\sigma^\prime| \hat T_{\mu\nu}^a(x) |p,\sigma\rangle
&=& \biggl[
2P_\mu P_\nu  \Bigl(
- {\ep^{\prime*}\cdot \ep} \, A^a_0 (t) 
 +{ {\ep^{\prime*}\cdot P} \, {\ep \cdot P} \over m^2}
 \, A^a_1(t) \Bigl)
\nonumber\\
&&+2\left[P_\mu(\ep^{\prime*}_{\nu} \,\ep\cdot P+\ep_{\nu}\,
\ep^{\prime*}\cdot P)
+P_\nu(\ep^{\prime*}_{\mu}\, \ep\cdot
P+\ep_{\mu} \,\ep^{\prime*}\cdot
P) \right] \, J^a (t)
\nonumber\\
&&+\frac12(\Delta_\mu \Delta_\nu-g_{\mu\nu}\Delta^2)
 \Bigl(
{\ep^{\prime*}\cdot \ep} \, D^a_0 (t)
+{ {\ep^{\prime*}\cdot P} \, {\ep \cdot P} \over m^2}  \, D^a_1(t)\Bigl)
\nonumber\\
&&+\Bigl[\frac12(\ep_{\mu}
\ep^{\prime*}_{\nu}+\ep^{\prime*}_{\mu}\ep_{\nu})\Delta^2
-(\ep^{\prime*}_{\mu}\Delta_\nu+\ep^{\prime*}_{\nu} \Delta_\mu)\,\ep\cdot P  
\nonumber\\
&& 
+(\ep_{\mu} \Delta_\nu+\ep_{\nu}
\Delta_\mu)\,\ep^{\prime*}\cdot P
-4g_{\mu\nu} \, {\ep^{\prime*}\cdot P} \, {\ep\cdot P} \Bigl] \, E^a(t)
\nonumber\\
&&
+\Bigl(\ep_{\mu}
\ep^{\prime*}_{\nu}+\ep^{\prime*}_{\mu}\ep_{\nu} - \frac{{\ep^{\prime*}\cdot \ep}}{2}\,g_{\mu\nu} \Bigl) \,{m^2} \, {\bar f}^a (t)
\nonumber\\
&& 
+g_{\mu\nu} \Bigl( {\ep^{\prime*}\cdot \ep} \, {m^2}\, {\bar c}^a_0(t)\, +  \, {\ep^{\prime*}\cdot P} \, {\ep \cdot P} \,{\bar c}^a_1(t)  \Bigl)  \biggr] \,e^{i(p^\prime-p)x} \ , 
\en
where $a=g,u,d,\dots$ and the polarization vectors $\ep^\prime_\mu=\ep_\mu(p^\prime,{\sigma^\prime})$, $\ep_{\mu}=\ep_{\mu}(p,\sigma)$, $\sigma=\pm 1,0$. 
 There are 9 GFFs for each quark flavour or gluon for a spin one  particles. 
The 6 quark and gluon from factors (FFs) $A^a_{0,1}$, $D^a_{0,1}$, $J^a$ and  $E^a(t)$ are individually EMT-conserving, and the other 3 FFs, ${\bar f}^a$ and ${\bar c}^a_{0,1}(t)$, are not\footnote{For the
particle with integer spin $J$ there are $(4 J+2)$ conserving and $(2J +1)$ non-conserving EMT FFs \cite{PSS}.}. 
As discussed in Ref.~\cite{Polyakov:2018zvc}, the all individual quark and gluon FFs depend on the renormalisation scale which we do not indicate for brevity. 
Due to EMT conservation, Eq.~(\ref{Eq:EMT-cons}), the constraint
$\sum_a\bar{f}^a(t)=0$ and $\sum_a\bar{c}^a_{0,1}(t)=0$ hold, and the total form factors
$A_{0,1}(t)$, $D_{0,1}(t)$, $J(t)$, $E(t)$ are renormalisation scale invariant where we
defined $A_{0}(t)\equiv \sum_a A^a_{0}(t)$% with $a=g,\,u,\,d,\,\dots$ 
and 
analogously for other form factors. Some of the notations for EMT FFs in the literature are listed in the Table~\ref{emtffs}. 
 The generalised form factors  in \cite{Cosyn:2018thq} and the reduced matrix elements in \cite{Hoodbhoy:1988am} are connected with the gravitational form factors as shown in Table \ref{emtffs}.  \\

\begin{table*}%[!htbp]
\caption{\label{emtffs} The notations of EMT FFs in the literature (the FFs in \cite{Cosyn:2018thq,Hoodbhoy:1988am} are not exactly EMT FFs ) and their values in free theory obtained by the Proca Lagrangian. 
%\footnote{%
	In Ref.~\cite{Holstein:2006ge} there is a sign mistake in the term correspoding to our $(\ep_{\mu} \Delta_\nu+\ep_{\nu} \Delta_\mu)\ep^{\prime*}\cdot P$ in $E^a(t)$'s coefficient. In Ref.~\cite{Taneja:2011sy}, the authors missed one term which should be corresponding to ${\bar c}^a_1$ term here. The result of Ref.~\cite{Cosyn:2019aio}, which appeared during the preparation of this paper, coincides with our result.
}
\begin{center}
	\begin{tabular}{c|ccccccccc}
	\hline
	\hline \\
	$\text{this work}$ & $A_0$ & $A_1$ & $D_0$ & $D_1$ & $J$ & $E$ & ${\bar f}$ & ${\bar c}_0$ & ${\bar c}_1$ \\ \hline \\
	$\text{free theory}$ & $1$ & $0$ & $1+4h$ & $0$ & $1$ & $1$ & $0$ & $0$ & $0$   \\ \hline \\
	$\text{Holstein~\cite{Holstein:2006ge}}$ & $F_1$ & $4F_5$ & $-2F_2$ & $8F_6$ & $F_3$ & $-2F_4$ & -- & -- & -- \\ \hline \\
	$\text{Abidin~\cite{Abidin:2008ku}}$ & $A$ & $-2E$ & $C$ & $-8F$ & $A+B$ & $D$ & -- & -- & -- \\ \hline \\
	$\text{Taneja~\cite{Taneja:2011sy}}$ & ${\cal G}_{1}$ & $-2{\cal G}_{2}$ & $-{\cal G}_{3}$ & $-2{\cal G}_{4}$ & $\half{\cal G}_{5}$ & $-\half{\cal G}_{6}$ &  ${1\over2m^2} {\cal G}_{7}$ &  ${1\over4m^2} {\cal G}_{7}+{\cal G}_{8}$ &  --\\ \hline \\
	$\text{Cosyn~\cite{Cosyn:2019aio}}$ & ${\cal G}_{1}$ & $-2{\cal G}_{2}$ & $-{\cal G}_{3}$ & $-2{\cal G}_{4}$ & $\half{\cal G}_{5}$ & $-\half{\cal G}_{6}$ &  $\half{\cal G}_{7}$ &  ${1\over4} {\cal G}_{7}+{\cal G}_{8}$ &  $-2 {\cal G}_{9}$ \\ \hline \hline \\
	$\text{Cosyn~\cite{Cosyn:2018thq} generalised form factors}$ & $A^a_{2,0}$ & $-2C^a_{2,0}$ & $-4F^a_2$ & $-8G^a_2$ & $\half B^a_{2,0}$ & $D^a_{2,1}$ &  $\sim E^a_{2,0}$ &  -- &  -- \\ \hline  \\
	$\text{Hoodbhoy~\cite{Hoodbhoy:1988am} reduced matrix elements}$ & $a_2-\frac13 d_2$ & -- & -- & -- & -- & -- &  $\sim d_2$ &  -- &  --  \\ 
	\hline
	\hline
	\end{tabular}
\end{center}
\end{table*}

\subsection{EMT form factors in free field theory }
In the free field theory, the massive spin one particles are described by the Proca Lagrangian, 
\begin{equation} \label{eq:Proca}
{\cal L}=-{1\over 4}U_{\mu\nu}U^{\mu\nu}+{1\over2}m^2 A_\mu A^\mu \ ,
\end{equation}
where $A_\mu$ is a massive vector field and the field tensor is 
\begin{equation} \label{eq:field_tensor}
U_{\mu\nu}=\partial_\mu A_\nu-\partial_\nu A_\mu \ . 
\end{equation}
The EMT corresponding to the Proca Lagrangian is given by 
\eq
{\hat T}^{\text{(Proca)}}_{\mu\nu} = - U_{\mu\rho} U_{\nu}^{\ \ \rho} - g_{\mu\nu} {\cal L} + m^2 A_\mu A_\nu \ .
\en

{ The action $S_{\rm grav}$ can be modified by adding a non-minimal term for interaction with the gravity:
\begin{equation} \label{eq:la_Imprv}
S_{\rm grav}=\int d^4 x \sqrt{-g} \bigg( -{1\over 4}U_{\mu\nu}U^{\mu\nu}+{1\over2}m^2 A_\mu A^\mu + \frac12 h R A^2 \bigg) %\label{eqn:la}
\end{equation} }
Here, $R$ is the Riemann scalar.
With this term added,  the EMT in the free field theory becomes: 
\eq
&&{\hat T}^{\rm improve}_{\mu\nu} = {\hat T}^{\text{(Proca)}}_{\mu\nu}+ \theta^{\rm improve}_{\mu\nu}   \ , \\
&&{\text{with, \;}} \theta^{\rm improve}_{\mu\nu} = -h
	(\partial_\mu\partial_\nu-g_{\mu\nu}\partial^2)\,A^2  \ .
\en
The value of the parameter $h$ depends on the physics problem one is considering.
With this improved EMT, one can obtain the free theory  values of the total FFs~\cite{Holstein:2006ge} as shown in Table.~\ref{emtffs}. 
\\

\section{The static EMT and stress tensor}
\label{sec:The-static-EMT-and-stress-tensor}

Before discussing the components of EMT in Eq.~(\ref{Eq:EMT-FFs-spin-1}), let us review the spin and quadrupole operators.  For particles with spin $S \ge 1$, the quadrupole operator is {the} $(2S+1)\times (2S+1)$
matrix:
\be \label{eq:quadrupole}
\hat Q^{ik}=\frac 12 \left( \hat S^{\ i} \hat S^{\ k}+\hat S^{\ k} \hat S^{\ i} -\frac 23 S(S+1) \delta^{ik} \right), \ (i,j,k=1,2,3) ,
\ee
which is expressed in terms of the spin operator $\hat S^{\ i}$.
%The matrix elements of ${\hat Q}^{kl}=\la S, \sigma^\prime \, | {\hat Q}^{kl} | S, \sigma\ra$. 
The spin operator can be expressed in terms of the SU(2) Clebsch-Gordan coefficients (in the spherical basis):
\be
\hat S^{\ \lambda}_{\sigma^\prime\sigma}= \sqrt{S(S+1)}\  C_{S \sigma 1\lambda}^{S \sigma^\prime} \ , \ (\lambda=0,\pm1. \ \sigma,\sigma^\prime=0,\cdots , \pm J) .
\ee
In spin one case, it is equivalent to  
\eq
\hat S^{\ i}_{\sigma^\prime \sigma}= i \ep^{ijk} \, \hat\ep_{\sigma}^{*\,j}  \hat\ep_{\sigma^\prime}^k \ , \ (i,j,k=1,2,3) ,
\en
where $\epsilon^\mu {(0,\sigma)}=(0,\hat\ep_\sigma)$ is the rest frame spin-1 polarization vectors, 
\begin{align} \label{eq:hat-ep}
\hat\ep_\pm  = \mp \frac{1}{\sqrt{2}}(1,\pm i,0), \
\hat\ep_0  = (0,0,1),
\end{align}

Applying the  boost operator $L(p)$ from the rest frame $k^\mu=(m,0,0,0)$ to any frame $p^\mu (={L}^\mu_{\; \nu}k^\nu)$, one has ~\cite{Weinberg:1995mt,Detmold:2017oqb}
\eq \label{eq:polarization_vector}
\ep^\mu ({p},\sigma) 
&=& \left( \frac{\vec{p}\cdot\hat\ep_\sigma}{m}, \hat\ep_\sigma + \frac{\vec{p}\cdot\hat\ep_\sigma}{m(m+{E})}\vec{p}\right), 
\en
where $\sigma=\{+,-,0\}$, $m$ and ${E}=\sqrt{|\vec{p}|^2+m^2}$ are the rest mass and energy of the state. 
\\ 

In the Breit frame, the initial(final) momentum $p^\mu$($p^{\prime\mu}$) has the relation $P^\mu=(p^\mu+p^{\prime\mu})/2=(E,0,0,0)$ and 
$\Delta^\mu=p^{\prime\mu}-p^\mu=(0,{\vec \Delta})$. So ${\vec p}=-{\vec p}^{\,{ }\prime}=-{\vec \Delta}/2$ and $p^0=p^{\prime\, 0}=E=\sqrt{m^2-t/4}$ with $t=\Delta^2$. 
In this frame, with the polarization vectors  Eq.~(\ref{eq:polarization_vector}), Eq.~(\ref{Eq:EMT-FFs-spin-1}) can be expressed as
\sub{
\label{eq:emt-breit-frame}
\eq
\langle p^\prime,\sigma^\prime| \hat T^{00}_a(0) |p,\sigma\rangle
\label{eq:thetauv00}
&=& 
2 m^2\mathcal{E}^a_{0}(t) \,  \delta_{\sigma^\prime \sigma}
+{\hat Q}^{kl} \, { \Delta}^k { \Delta}^l \, \mathcal{E}^a_2(t)  \ , 
\\ [0.5em]
\langle p^\prime,\sigma^\prime| \hat T^{0j}_a(0) |p,\sigma\rangle
\label{eq:thetauv0j} 
&=&
 {i \ep^{jkl} {\hat S}^k_{\sigma^\prime \sigma}}\Delta^l \, m\, \mathcal{J}^a (t)\ , 
\\ [0.5em]
\langle p^\prime,\sigma^\prime| \hat T^{ij}_a(0) |p,\sigma\rangle
\label{eq:thetauvij} 
&=&
\frac12(\Delta^i \Delta^j-\delta^{ij}\vec\Delta^2) 
{\mathcal D}^a_0(t) \,  \delta_{\sigma^\prime \sigma}
%\nonumber\\
%&&
+ \left( 
\Delta^j \Delta^k {\hat Q}^{ik}      
+ \Delta^i \Delta^k {\hat Q}^{jk} 
- \vec\Delta^2 {\hat Q}^{ij}
-\delta^{ij} { \Delta}^k { \Delta}^l {\hat Q}^{kl} \right) \, {\mathcal D}^a_2(t)
\nonumber\\
&&
+{ 1 \over 2m^2} (\Delta^i \Delta^j-\delta^{ij}\vec\Delta^2) { \Delta}^k { \Delta}^l {\hat Q}^{kl} \,
{\mathcal D}^a_3(t)
\nonumber\\
&&
+ \biggl[ \delta^{ij} \delta_{\sigma^\prime \sigma} \Bigl( {m^2\over6}+{t \over 12} \Bigr) + {1\over6} \Delta^i \Delta^j \delta_{\sigma^\prime \sigma} - 2m^2{\hat Q}^{ij} 
-{m\over2(m+E)} \bigl( \Delta^i \Delta^k {\hat Q}^{kj} + \Delta^j \Delta^k {\hat Q}^{ki}\bigr) 
\nonumber\\
&&
+{1\over4} \Bigl( \delta^{ij} -{\Delta^i \Delta^j \over2(m+E)^2} \Bigr) \Delta^k \Delta^l {\hat Q}^{kl} \biggr]\, {\bar f}^a(t) 
\nonumber\\
&&
+\delta^{ij} \biggl\{ \Bigl[ \delta_{\sigma^\prime \sigma} \Bigl( {t\over6}-m^2 \Bigr)+ {1\over2} \Delta^k \Delta^l {\hat Q}^{kl} \Bigr] \, {\bar c}^a_0(t) 
+ \frac14\Bigl( 1-{t\over4m^2} \Bigr)\Bigl( {t\over3} \delta_{\sigma^\prime \sigma} +\Delta^k \Delta^l {\hat Q}^{kl} \Bigr)\,  {\bar c}^a_1(t)\biggr\} , \
\en }
where ${\hat Q}^{kl}=\la S, \sigma^\prime \, | {\hat Q}^{kl} | S, \sigma\ra$ are the matrix elements of the quadrupole operator and 
\sub{
\label{eq:gffs}
\eq 
\label{eq:E0}
\mathcal{E}^a_{0}(t)
&=& 
 A^a_0(t) + {1\over4} {\bar f}^a(t) - {1\over2} {\bar c}^a_0(t) 
\nonumber \\
&&
+{t\over {12}m^2} \Bigl[ -{5} A^a_0(t)  +3 D^a_0(t)+ 4J^a(t) -2E^a(t) +A^a_1(t) +{1\over2} {\bar f}^a(t) + {\bar c}^a_0(t)+ \frac12 {\bar c}^a_1(t) \Bigr]
\nonumber \\
&&
-{t^2\over {24}m^4} \Bigl[ -A^a_0(t)  + D^a_0(t) + {2}J^a(t) - 2E^a(t) + A^a_1(t) +  {1\over2}D^a_1(t) + \frac14 {\bar c}^a_1(t) \Bigr]
+{t^3\over {192}m^6}  \Bigl[ A^a_1(t) +  D^a_1(t)  \Bigl] , \ \\ [0.7em]
\label{eq:E2}
\mathcal{E}^a_{2}(t) 
&=& 
-A^a_0(t) + {2}J^a(t)  -E^a(t)  + \frac12A^a_1(t) + \frac14 {\bar f}^a(t)+\frac12 {\bar c}^a_0(t)+ \frac14 {\bar c}^a_1(t)
\nonumber \\
&& 
- {t\over 4m^2} \Bigl[ -A^a_0(t)  + D^a_0(t) + 2J^a(t) - 2E^a(t) + A^a_1(t) +  {1\over2}D^a_1(t) + \frac{1}{4} {\bar c}^a_1(t) \Bigr]
\nonumber \\
&& + {t^2\over 32m^4}  \Bigl[ A^a_1(t) +  D^a_1(t)  \Bigl]  \ , \\ [0.5em]
\label{eq:J}
\mathcal{J}^a(t)&=& 
 J^a(t) + {1\over2} {\bar f}^a(t) -{t\over4m^2} \Bigl( J^a(t)   -  E^a(t) \Bigl) \ . \\
\label{eq:D0}
\mathcal{D}^a_0(t)&=& 
-D^a_0(t)+{4\over3} E^a(t) +{t\over12m^2} \Bigl[ 2D^a_0(t)-2 E^a(t)+ D^a_1(t) \Bigl] -  {t^2\over48m^4}  D^a_1(t) \ , \\
\label{eq:D2}
{\mathcal D}^a_2(t)&=& -E^a(t) \ , \\
\label{eq:D3}
{\mathcal D}^a_3(t)&=&  \frac14 \Bigl[ 2D^a_0(t)-2 E^a(t)+ D^a_1(t) \Bigl] - {t \over 16m^2}   D^a_1(t) \ . 
\en }
The details for obtaining  Eq.~(\ref{eq:emt-breit-frame}) and (\ref{eq:gffs}) are shown in Appendix~\ref{sec:apdix-gffs}. 

\begin{table*}[t]%!htbp]
\caption{\label{lovalues} The free theory values of the total EMT FFs. }
\begin{center}
	\begin{tabular}{c|cccccc}
	\hline
	\hline \\
	{\text{EMT FFs}}  & $\mathcal{E}_{0}(t)$ & $\mathcal{E}_{2}(t)$ & $\mathcal{J}(t)$ & $\mathcal{D}_{0}(t)$ & $\mathcal{D}_{2}(t)$ & $\mathcal{D}_{3}(t)$  \\ \hline \\
	{\text{free theory}}  & $1$ & 0 & $1$ & $\frac13-4h$ & -1 & 0 \\ \hline 
	\hline
	\end{tabular}
\end{center}
\end{table*}

Due to the constraint $\sum_a\bar{f}^a(t)=0$ and $\sum_a\bar{c}^a_{0,1}(t)=0$, the total quark $+$ gluon EMT drop $\bar f^a$ and $\bar{c}^a_{0,1}(t)$ terms, so do $\mathcal{E}_{0,2}$, $\mathcal{J}$ and $\mathcal{D}_{0,2,3}(t)$. The free theory values of the total EMT FFs are listed in Table~\ref{lovalues}.  The D-term is defined as 
\eq
D \equiv \mathcal{D}_{0}(0)= \frac13 - 4 h \ .
\en

Following Ref.~\cite{Polyakov:2002yz}, the static EMT $T^{\mu\nu}(\vec r, \sigma^\prime,\sigma)$ of the spin-1 particle is defined  by Fourier tranforming the EMT in Eq.~(\ref{eq:thetauv00},\ref{eq:thetauv0j},\ref{eq:thetauvij}) with respect to $\vec \Delta$ as 
\eq \label{eq:static_EMT}
T^{\mu\nu}_a(\vec r, \sigma^\prime,\sigma) &=& \int {d^3 \Delta \over 2E(2\pi)^3 } e^{-i \vec \Delta \cdot \vec r}  \langle p^\prime, \sigma^\prime \, |{\hat T}^{\mu\nu}_a(0)|p,\sigma \rangle \ ,
\en
where $r = \abs{\vec r \,}$. 

\subsection{$T^{00}$: Energy density}
\label{subsec:Energy-density}

Due to the presence of the EMT-nonconserving terms ${\bar f}^a$ and ${\bar c}^a_{0,1}$, the energy density $T^{00} ( \vec r, \sigma^\prime,\sigma)$ can only be defined for the total system. 
The multipole expansion of the energy density is defined as~\cite{Polyakov:2018rew}
\eq
T^{00} ( \vec r, \sigma^\prime,\sigma) 
&=& 
\int {d^3 \Delta \over 2E (2\pi)^3 } e^{-i \vec \Delta \cdot \vec r} 
\langle p^\prime, \sigma^\prime \, |{\hat T}^{00}(0)|p,\sigma \rangle 
\\ 
&=& \label{eq:thetauv00BA}
\varepsilon_0(r) \, \delta_{\sigma^\prime\sigma}
+\varepsilon_2(r)\, {\hat Q}^{ij} \, Y^{ij}_2
\ , 
\en
where 
\eq
&&\varepsilon_0(r) = 2m^2{\tilde {\mathcal E}}_{0}(r) \ , \
\varepsilon_2(r) = -{r} {d\over dr} {1\over r} {d\over dr} {\tilde {\mathcal E}}_{2}(r) \ , \\
\label{eq:ft2E}
&&\text{with:}\quad   {\tilde {\mathcal E}}_{0,2}(r) =\int{d^3\Delta\over 2E(2\pi)^3} e^{-i\vec\Delta\cdot\vec r} {\mathcal E}_{0,2}(t)
\en 
(the defination of Eq.~(\ref{eq:ft2E}) is used for other FFs in the following), 
and the irreducible (symmetric and traceless) tensor  of $n$-th rank are~\cite{Polyakov:2018rew}:
\eq
Y_{n}^{i_1 i_2 ... i_n} = \frac{(-1)^n}{(2 n-1)!!} r^{n+1} \partial^{i_1}...\partial^{i_n} \frac{1}{r}, 
%\ee
%\be
\quad \mbox{i.e.} \quad
Y_0=1,\ Y_1^{i}=\frac{r^{i}}{r}, \ Y_2^{ik}=\frac{r^{i} r^{k}}{r^2}-\frac13 \delta^{ik}, \ {\rm etc.}  \ .
\en
Notes there are obvious relations $\delta^{i_l i_m}Y_{n}^{i_1 i_2 ... i_n}=0$ and $\int d\Omega\, Y_2^{ik}=0$. 

In Ref.~\cite{Polyakov:2018rew}, more general tensor quantities are introduced for a particle of arbitrary spin:
\be
\label{eq:massmulti}
M_n^{k_1\ldots k_n}=\int d^3 {\bf r}\   r^{n}\ Y_n^{k_1 \ldots k_n} \ T^{00} ( \vec r \,) ,
\ee   
which correspond {to} $2^n$-multipoles of the energy distribution. Here $T^{00}({\vec r})=T^{00}({\vec r},\sigma,\sigma)$. Note, that only even $n$ are allowed by the $P$-parity conservation. Obviously ,
\eq
M_0=m A_0(0) =m \ , 
\en
which gives the normalisation
\eq \label{eq:normalisation-A0}
 A_0(0) =\sum_a A^a_0(t) =1 \ ,
\en

The function $\varepsilon_2(r)$  gives the quadrupole distribution of the energy inside the particles and describe the deviation of the hadron's shape from the spherical one for
the hadrons with spin larger than 1/2. Obviously it satisfies the condition $\int d^3 r \varepsilon_2(r) =0$. For free spin-1 particle one obtains (see Table~\ref{lovalues}) that the
quadrupole energy distribution is zero. Intuitively clear result.  

\subsection{$T^{0j}$: Spin distribution}
\label{subsec:spin-distribution}

The $0k$-components of the EMT are related to the spatial distribution of the spin. The angular momentum operator in QCD is defined
according to the generators of Lorentz transformation\cite{Ji:1996ek},
\begin{equation} \label{eq:amo}
     J^i = {1\over 2}\epsilon^{ijk} 
      \int d^3x M^{0jk} \ , 
\end{equation}
where $M^{0ij}$ is the angular momentum density, expressible
in terms of the energy-momentum tensor $T^{\mu\nu}$ through
\begin{equation}
      M^{\alpha\mu\nu} = T^{\alpha\nu} x^\mu - 
      T^{\alpha\mu} x^\nu \ . 
\end{equation}
From Eq.~(\ref{eq:thetauv0j}), one gets
\eq
T^{0j} (\vec r, \sigma^\prime,\sigma) 
&=&  \label{Eq:EMT-Breit-T0k}
\int {d^3 \Delta \over 2E (2\pi)^3 } e^{-i \vec \Delta \cdot \vec r} 
\langle p^\prime, \sigma^\prime \, |{\hat T}_a^{0j}(0)|p,\sigma \rangle \ .
\en

According to Eq.~(\ref{eq:amo}), define the individual contributions of quarks and gluons to the particle spin ($J=1$) as~\cite{Ji:1996ek,Lorce:2017wkb,Polyakov:2018zvc}, 
\eq \label{Eq:static-EMT-T0k-1}
	J_a^i(\vec{r},{\sigma^\prime},\sigma)=\epsilon^{ijk}r^jT^{0k}(\vec{r},\sigma^\prime,\sigma)\;.
\en
Inserting the expression (\ref{Eq:EMT-Breit-T0k}) (with Eq. (\ref{eq:thetauv0j}))
into Eq.~(\ref{Eq:static-EMT-T0k-1}) yields:
\eq \label{Eq:static-EMT-T0k-2}
	J_a^i(\vec{r},\sigma^\prime,\sigma) 
={\hat S}^j_{\sigma^\prime\sigma}  \int {d^3 \Delta \over  (2\pi)^3 } e^{-i \vec \Delta \cdot \vec r}  \bigg[ \Big(  \mathcal{\bar J}^a(t) +\frac23 t {d \mathcal{\bar  J}^a(t) \over dt}  \Big) \delta^{ij} + \Big( \Delta^i\Delta^j - \frac13 \vec \Delta^2 \delta^{ij}  \Big) {d \mathcal{\bar J}^a(t) \over dt}  \bigg]\;,
\en
with $\mathcal{\bar J}^a(t)=\frac{m}{E} \mathcal{J}^a(t)$. Above equation has the form very similar to that for spin-1/2 particle ~\cite{Lorce:2017wkb,Polyakov:2018zvc}.
Note that the form factor $\mathcal{J}^a(t)$ contains the EMT non-conserving form factor $\bar f^a(t)$. In the case of spin-0 and spin-1/2 the non-conserving form factors do not enter the 
spatial spin distribution.

Summing over quarks and gluons and integrating over the space yields
\eq \label{eq:normalisation-J}
\sum_a \int d^3 r \, J_a^i(\vec{r},{\sigma^\prime},\sigma) = {\hat S}^i_{\sigma^\prime\sigma} {J(0)}={\hat S}^i_{\sigma^\prime\sigma} \ .  \  
\en 
where the individual contributions $J^a(0)$ add up to $\sum_a J^a(0)=J(0)$ which satisfies the normalisation condition $J(0)=J=1$. 
Obviously the free theory value $J^{\rm free\ theory}(t)=1$ in Table~\ref{emtffs} satisfies this relation. \\

\subsection{$T^{ij}$ Stress tensor}
\label{subsec:Stress-tensor}

In the sprit of Ref.~\cite{Polyakov:2018rew}, the stress tensor defined by the $ij$-components of EMT in Eq.~(\ref{eq:thetauvij}), can be written generically to the quadrupole order as:
\eq \label{eq:thetauvij_2}
T^{ij} (\vec r, \sigma^\prime,\sigma ) 
&=& 
\int {d^3 \Delta \over 2E (2\pi)^3 } e^{i \vec \Delta \cdot \vec r} 
\langle p^\prime, \sigma^\prime \, |{\hat T}_a^{ij}(0)|p,\sigma \rangle 
\nonumber \\ 
&=& \label{eq:thetauvij_BA_FT_0}
p_0(r) \delta^{ij} \delta_{\sigma^\prime \sigma} 
+s_0(r)Y_2^{ij}\delta_{\sigma^\prime \sigma} 
+ p_2(r) \hat{Q}^{ij} +2 s_2(r) 
\left[\hat{Q}^{ip}Y_{2}^{pj}+\hat Q^{jp}Y_{2}^{pi} -\delta^{ij} \hat Q^{pq}Y_{2}^{pq}  \right]
\nonumber\\
&&
- {1\over m^2} {\hat Q}^{kl} \partial_k \partial_l  
\bigl[ p_3(r) \delta^{ij} +s_3(r)Y_2^{ij} \bigr]
  \ ,
\en
where the (quadrupole) pressure and shear forces functions
\sub{
\eq 
\label{eq:pressure_force_0}
p_0(r) &=&  {1\over 3} \ \partial^2\ {\tilde {\cal D}}_0(r) \ , \  
s_0(r) = -{1\over 2} r{d\over dr} {1\over r}{d\over dr} \, {\tilde {\mathcal D}}_0(r) \ ,  \\
\label{eq:pressure_force_2}
p_2(r) &=& {1\over 3}\ \partial^2\ {\tilde {\mathcal D}}_2(r)   \ , \
s_2(r) = -{1\over 2} r{d\over dr}  {1\over r}{d\over dr} {\tilde {\mathcal D}}_2(r)   \ ,  \\
\label{eq:pressure_force_3}
p_3(r) &=& {1\over 3}\ \partial^2\ {\tilde {\mathcal D}}_3(r)\ ,  \
s_3(r) = -{1\over 2} r{d\over dr}  {1\over r}{d\over dr} {\tilde {\mathcal D}}_3(r)   \ .
\en }
where $\partial^2=\frac{1}{r^2}\frac{d}{dr} r^2\frac{d}{dr}$ is the radial part of 3D Laplace operator. 

Comparing with Ref.~\cite{Polyakov:2018rew}, we get two additional terms of quadrupole order $n=2$, which are $p_3(r)$ and $s_3(r)$ terms. 
The EMT conservation, $\partial_\mu {\hat T}^{\mu\nu}(x) =0$, implies the equilibrium equation for the static stress tensor
\eq \label{eq:conservation_0}
\partial_i T^{ij}(\vec r, \sigma^\prime, \sigma) = 0 \ .  
\en

For each of the first two quadrupole orders, it is easy to check that Eq.~(\ref{eq:pressure_force_0}),(\ref{eq:pressure_force_2}),(\ref{eq:pressure_force_3}) satisfy the differential equations  
\eq \label{eq:conservation_2}
{2\over3}  s'_{n}(r) + 2{s_{n}(r) \over r} +  p'_{n}(r) =0 \ , {\rm with }\  n=0,2,3, \ 
\en
which guarantee the general stability condition of Eq.~(\ref{eq:conservation_0}). 

Another three obvious relations, 
\eq
\int d^3 {\bf r} \, p_n(r) =\frac{1}{3}\int d^3 {\bf r}\ \partial^2\   {\mathcal D}_n (r)=0 \ , {\rm with }\  n=0,2,3, \ 
\en
which shows how the internal forces balance inside a composed particle. It is a consequence of the EMT conservation, known as the von Laue condition~\cite{Polyakov:2018zvc,Lorce:2018egm}.  
We also note that the multipole pressure and shear forces distributions ($p_n(r), s_n(r)$) satisfy the same stability equation (\ref{eq:conservation_2}) as the distributions in spherically symmetric
case of spin-0 and spin-1/2 particles. Therefore all stability relations discussed  in \cite{Polyakov:2018zvc,Lorce:2018egm} are valid also for non-spherically case of particles with higher spins.

\subsection{EMT-nonconserving terms }

The EMT-nonconserving terms in Eq. (\ref{Eq:EMT-FFs-spin-1}) violate the EMT conservation {     $\partial^\mu {\hat T}_{\mu\nu}(x) =0$ as
\eq 
\langle p^\prime,\sigma^\prime| {\partial^\mu} \hat T_{\mu\nu}^a(x) |p,\sigma\rangle
&=& i \Delta^\mu \, \langle p^\prime,\sigma^\prime|  \hat T_{\mu\nu}^a(x) |p,\sigma\rangle \nonumber \\
&=&   \label{Eq:EMT-FFs-spin-2}
 ie^{i\Delta x}  \biggl[
\Bigl(\ep \cdot \Delta
\ep^{\prime*}_{\nu}+\ep^{\prime*} \cdot \Delta \ep_{\nu} - \frac{{\ep^{\prime*}\cdot \ep}}{2}\,  \Delta_{\nu} \Bigl) \,{m^2} \, {\bar f}^a (t)
\nonumber\\
&& 
+  \Delta_{\nu} \Bigl( {\ep^{\prime*}\cdot \ep} \, {m^2}\, {\bar c}^a_0(t)\, +  \, {\ep^{\prime*}\cdot P} \, {\ep \cdot P} \,{\bar c}^a_1(t)  \Bigl)  \biggr] \ , 
\en }
In Breit frame, the $0$-component of Eq. (\ref{Eq:EMT-FFs-spin-2}) is 
{
\eq
\langle p^\prime,\sigma^\prime| {\partial^\mu} \hat T_{\mu0}^a(x) |p,\sigma\rangle
&=&   \label{Eq:EMT-FFs-spin-20}
 ie^{i\Delta x} 
\Bigl(\ep \cdot \Delta
\ep^{\prime*}_{0}+\ep^{\prime*} \cdot \Delta \ep_{0}  \Bigl) \,{m^2} \, {\bar f}^a (t)  \nonumber \\
&=&  0 \ , 
\en }
and the $j$-component of Eq. (\ref{Eq:EMT-FFs-spin-2}) is
{    
\eq
\langle p^\prime,\sigma^\prime| {\partial^\mu} \hat T_{\mu j}^a(x) |p,\sigma\rangle
&=&  \label{Eq:EMT-FFs-spin-2i}
 ie^{i\Delta x}  \biggl[
\Bigl(\ep \cdot \Delta
\ep^{\prime*}_{i}+\ep^{\prime*} \cdot \Delta \ep_{i} - \frac{{\ep^{\prime*}\cdot \ep}}{2}\,  \Delta_{i} \Bigl) \,{m^2} \, {\bar f}^a (t)
\nonumber\\
&& 
+  \Delta_{i} \Bigl( {\ep^{\prime*}\cdot \ep} \, {m^2}\, {\bar c}^a_0(t)\, +  \, {\ep^{\prime*}\cdot P} \, {\ep \cdot P} \,{\bar c}^a_1(t)  \Bigl)  \biggr] \nonumber \\ [0.5em]
&=&  % \label{Eq:EMT-FFs-spin-2i}
 ie^{i\Delta x}  \biggl\{
 \Delta^j \delta_{\sigma^\prime \sigma} m^2 \bigg[   { {\bar f}^a(t) \over 6 }- {\bar c}_0^a(t) + {t\over 12m^2} \big[ - { {\bar f}^a(t) } + 2{ {\bar c}_0^a(t) }+ {{\bar c}_1^a(t) } \big]
 -{t^2\over 48m^4} {\bar c}_1^a(t) \bigg]
 \nonumber\\
&&
- 2mE\Delta^i \, {\hat Q}^{ij}  {\bar f}^a(t) \, 
%\nonumber\\
%&&
+  \Delta^j \Delta^k \Delta^i {\hat Q}^{ki} \biggl[ \frac12 {\bar c}_0^a(t) + \frac14 {\bar c}_1^a(t) + {t \over16(m+E)^2} \, {\bar f}^a(t) - {t\over 16m^2} {\bar c}_1^a(t)  \biggr]
\biggr\} 
%\nonumber \\[0.5em]
 \ .
\en }
The stability equation for the quark part of the stress tensor has the form:
\be
%\nonumber
\frac{\partial T_{ik}^q({\bf r})}{\partial r^k} +f^i({\bf r})=0. 
\ee
This equation can be interpreted (see e.g discussion in \cite{Polyakov:2018exb}) as the equilibrium 
equation for quark internal stress and external force (per unit of the volume) $f^i({\bf r})$ acting on quark subsystem from the side of the gluons.
From Eq.~(\ref{Eq:EMT-FFs-spin-2i}) one sees that the corresponding force depends on the polarisation of the spin-1 particle through the quadrupole 
spin operators ${\hat Q}^{ij}$.

\section{Sum rules: GPD and GFF}
\label{sec:Sum-rules-GPD-and-GFF}

By considering the Mellin moments of the vector generalised parton distributions (GPDs)~\cite{Berger:2001zb}, the sum rules between the GPDs and EMT FFs are found in Ref.~\cite{Abidin:2008ku,Taneja:2011sy}. The sum rules in Ref.~\cite{Abidin:2008ku} contain only for conserving EMT FFs ( 6 out total 9 FFs).
 In recent paper \cite{Cosyn:2018thq}, the polynomiality sum rules for all leading-twist quark and gluon generalised parton distributions of spin-1 targets are given. 
The generalised form factors in the polynomiality condition for GPDs of spin-1 particles 
in \cite{Cosyn:2018thq}  are connected to the gravitational form factors as shown in Table \ref{emtffs}.

The quark and gluon vector GPDs are introduced in Ref.~\cite{Berger:2001zb} for deuteron as:
\sub{
\label{eq:GPDs}
\ba
\label{eq:qGPDs}
\lefteqn{ \frac{1}{2} \int \frac{d z^-}{2\pi}\, e^{ix P^+ z^-}
  \langle  p', \sigma^\prime |\, \bar{\psi}_q(-\half z)\, \gamma^+ \, \psi_q(\half z)\,
  \,| p, \sigma \rangle \Big|_{z^+ = 0,\, {\bf z}_\perp=0 } }
\nonumber \\ 
&=& -(\ep'^*\cdot\ep)H_1^q +\frac{\ep^+(\ep'^*\cdot P)+\ep'^{*+}(\ep\cdot P)}{P^+}H_2^q
-\frac{2(\ep\cdot P)(\ep'^*\cdot P)}{M^2}H_3^q
\nonumber \\ &&
+\frac{\ep^+(\ep'^*\cdot P)-\ep'^{*+}(\ep\cdot P)}{P^+}H_4^q
+\left\{ M^2\frac{\ep^+ \ep'^{*+}}{(P^+)^2} + \frac{1}{3}(\ep'^*\cdot\ep) \right\} H_5^q \ , 
\\
\label{eq:gGPDs}
\lefteqn{ \frac{1}{P^+} \int \frac{d  z^-}{2\pi}\, e^{ix  P^+ z^-}
  \langle  p', \sigma^\prime |\, F^{b,+\eta}(-\half z)\,{F^{b,}}_{\eta}{ }^+(\half z)
  \,| p, \sigma \rangle \Big|_{z^+ = 0,\, {\bf z}_\perp=0 } }
\nonumber \\ 
&=& -(\ep'^*\cdot\ep)H_1^g +\frac{\ep^+(\ep'^*\cdot P)+\ep'^{*+}(\ep\cdot P)}{P^+}H_2^g
-\frac{2(\ep\cdot P)(\ep'^*\cdot P)}{M^2}H_3^g
\nonumber \\ &&
+\frac{\ep^+(\ep'^*\cdot P)-\ep'^{*+}(\ep\cdot P)}{P^+}H_4^g
+\left\{ M^2\frac{\ep^+ \ep'^{*+}}{(P^+)^2} + \frac{1}{3}(\ep'^*\cdot\ep) \right\} H_5^g \ , 
\ea}
where $q=u,d,s,\dots$.  Integrating over $x$ of Eq.~(\ref{eq:qGPDs}), 
one gets the conventional form factor decomposition of the vector current for a spin-1 particle, 
\eq\label{eq:Imm}
%J_{\sigma^\prime\sigma}^\mu &=& 
\langle  p',  {\sigma^\prime} |\, \bar{\psi}_q(0)\, \gamma^\mu \, \psi_q(0)\,| p, \sigma \rangle
%\nonumber \\ %&& {}
&=&  -2\Big( \ep'^{*} \cdot \ep \, G_1^q(t) 
+ 2G_3^q(t)  \frac{ \ep'^{*} \cdot P \,  \ep  \cdot P}{m^2} \Big) P^\mu
%\nonumber \\ && {}
+ 2G_2^q(t) \left( \ep^\mu \ep'^{*}  \cdot P  + \ep'^{*\mu} \ep \cdot P \right)  \ . 
\en
So, for the quark GPDs, one has~\cite{Berger:2001zb}
\sub{
\label{eq:sum-rules-GPD-FF}
\eq
\int_{-1}^{1} dx H_i^q (x,\xi,t) &=& G_i^q(t) \quad (i=1,2,3) \ ,  \\
\label{eq:sum-rules-GPD-FF-45}
\int_{-1}^{1} dx H_i^q (x,\xi,t) &=& 0  \quad (i=4,5) \ .
\en}
The charge, magnetic, and quadruploe form factors can be expressed in terms of $G_i=\sum_q G^q_i$ as ( $\eta=-t/4m^2$)
\sub{
\label{eq:Gcmq}
\begin{eqnarray}
G_C(t)&=&G_1(t) + \frac{2}{3}{\eta} G_Q(t) \ ,  \\
G_M(t)&=&G_2(t) \ ,  \\
G_Q(t)&=&G_1(t) - G_2(t) + (1+\eta)G_3(t)\ , \
\end{eqnarray} }
normalised by the charge $G_C(0)=1$, magnetic moment $G_M(0)=\mu_{S=1}/(2m)$, and quadrupole moment $G_Q(0)=Q_{S=1}/m^2$.   

The $++$ components of Eq.~(\ref{eq:EMT-QCD}) are, 
\sub{
\label{eq:EMT-QCD-pp}
\ba
	T^{++}_q (x) &=& \frac{1}{2}\overline{\psi}_q \biggl(
	-i\overset{ \leftarrow}{\cal D}{ }^+\gamma^+
	+i\overset{\rightarrow}{\cal D}{ }^+\gamma^+ \biggr)\psi_q (x) \ , \\
	T^{++}_g (x) &=& F^{b,+\eta}\,{F^{b,}}_{\eta}{ }^+(x) \ , 
\ea}
or,
\sub{
\label{eq:EMT-QCD-pp-2}
\ba
	T^{++}_q (0) &=&(P^+)^2 \int xdx \,  \int \frac{d  z^-}{2\pi}\, e^{ix  P^+ z^-}  \biggl[ \overline{\psi}_q(-\half z) \gamma^+ \psi_q (\half z) \biggr]_{z^+ = 0,\, {\bf z}_\perp=0 } \ , \nonumber  \\
	&=& \biggl[
2(P^+)^2   \Bigl(
- {\ep^{\prime*}\cdot \ep} \, A^q_0 (t) 
 +{ {\ep^{\prime*}\cdot P} \, {\ep \cdot P} \over m^2}
 \, A^q_1(t) \Bigl)
\nonumber\\
&&+\frac12( \Delta^+)^2
 \Bigl(
{\ep^{\prime*}\cdot \ep} \, D^q_0 (t)
+{ {\ep^{\prime*}\cdot P} \, {\ep \cdot P} \over m^2}  \, D^q_1(t)\Bigl)
\nonumber\\
&&+4P^+ (\ep^{\prime*+} \,\ep\cdot P+\ep^{+}\,
\ep^{\prime*}\cdot P) \, J^q (t)
\nonumber\\
&&+\Bigl[ \ep^{+}
\ep^{\prime*+} \Delta^2
-2\ep^{\prime*+}\Delta^+ \,\ep\cdot P  
+2\ep^{+} \Delta^+  \,\ep^{\prime*}\cdot P
  \Bigl] \, E^q(t)
\nonumber\\
&&
+2\ep^+ 
\ep^{\prime*+}  \,{m^2} \, {\bar f}^q (t) \ , \\[0.7em]
	T^{++}_g (0) &=& P^+ \int dx \,  \int \frac{d  z^-}{2\pi}\, e^{ix  P^+ z^-}  
\biggl[ F^{b,+\eta}(-\half z)\,{F^{b,}}_{\eta}{ }^+(\half z)  \biggr]_{z^+ = 0,\, {\bf z}_\perp=0 }  \ .
\ea}
where $T^{++}_g (0)$ is similar with $T^{++}_q (0)$. 
Compare Eq.~(\ref{eq:EMT-QCD-pp-2}) with (\ref{eq:GPDs}), we get the  polynomiality property of vector GPDs as
\sub{
\label{eq:sum-rule-q}
\eq
\label{eq:sum-rule-q-1}
\int_{-1}^1 xdx H^q_1 (x,\xi,t) &=& A^q_0(t)-\xi^2 D^q_0(t)+ {t\over 6m^2} E^q(t) + \frac13{\bar f}^q(t) \ ,   \\
\int_{-1}^1 xdx H^q_2 (x,\xi,t) &=& 2J^q(t) \ ,    \label{eq:spin-1-ji-sum-rule-q}  \\
\int_{-1}^1 xdx H^q_3 (x,\xi,t) &=& -\frac12 \left[ A^q_1(t) +\xi^2 D^q_1 (t) \right] \ ,   \\
\int_{-1}^1 xdx H^q_4 (x,\xi,t) &=& -2 \xi E^q(t) \ ,   \\
\int_{-1}^1 xdx H^q_5 (x,\xi,t) &=& {t\over 2m^2} E^q(t) + {\bar f}^q(t) \ ,   
\en }
for the quark parts and 
\sub{
\label{eq:sum-rule-g}
\eq
\int_{-1}^1 dx H^g_1 (x,\xi,t) &=& A^g_0(t)-\xi^2 D^g_0(t)+ {t\over 6m^2} E^q(t) + \frac13{\bar f}^g(t) \ ,   \\
\int_{-1}^1 dx H^g_2 (x,\xi,t) &=& 2J^g(t) \ ,   \label{eq:spin-1-ji-sum-rule-g} \\
\int_{-1}^1 dx H^g_3 (x,\xi,t) &=& -\frac12 \left[ A^g_1(t) +\xi^2 D^g_1 (t) \right] \ ,   \\
\int_{-1}^1 dx H^g_4 (x,\xi,t) &=& -2 \xi E^g(t) \ ,   \\
\int_{-1}^1 dx H^g_5 (x,\xi,t) &=& {t\over 2m^2} E^g(t) + {\bar f}^g(t) \ ,   
\en }
for the gluon part. Eq. (\ref{eq:spin-1-ji-sum-rule-q}) and (\ref{eq:spin-1-ji-sum-rule-g}) give the spin-1 version of the X. Ji sum
rule \cite{Abidin:2008ku,Cosyn:2019aio,Ji:1996ek}.  Note that, in $H^{q}_{1,5}$ ($H^{g}_{1,5}$) it contains the EMT-nonconserving GFFs ${\bar f}^q$ (${\bar f}^g$). Thus  it is useful to rewrite them as 
\sub{
\label{eq:sum-rule-A-0}
\eq
\label{eq:sum-rule-A-0-q}
\int_{-1}^1 xdx \left[ H^q_1 (x,0,t) -\frac13 H^q_5 (x,0,t) \right] &=& A^q_0(t) \ , \\
\int_{-1}^1 dx \left[ H^g_1 (x,0,t) -\frac13 H^g_5 (x,0,t) \right] &=& A^g_0(t) \ .
\en}
In the GPD approach, the normalisation for $A_0$ in Eq.~(\ref{eq:normalisation-A0}) is  the energy-momentum sum rule, 
\eq \label{eq:momentum-sum-rule}
\int_{-1}^1 dx \left[ \sum_q xH^q_1 (x,0,0)+H^g_1 (x,0,0) -\frac13 \bigg(  \sum_q  xH^q_5 (x,0,0) + H^g_5 (x,0,0) \bigg) \right]=1 \ , 
\en
and the normalisation for $J$ in Eq. (\ref{eq:normalisation-J}) gives the sum rule,
\eq \label{eq:spin-sum-rule}
\frac12 \int_{-1}^1 dx \, \left[ \sum_q  x H^q_2 (x,0,0) + H^g_2 (x,0,0) \right]=1 \ .
\en 

As discussed in Ref. \cite{Berger:2001zb}, the forward limit ($\xi=t=0$) of GPDs $H_{1}$ and $H_{5}$ can give the parton distributions functions (PDFs) and therefore $H_{1,5}$ are respectively related to the DIS structure functions $F_1(x)$ and $b_1(x)$ which are worked out by Hoodbhoy {\it et al.} \cite{Hoodbhoy:1988am} for spin one targets. Thus, in terms of PDFs, the normalisation for $A_0$ in Eq. (\ref{eq:momentum-sum-rule}) corresponds to the energy-momentum conservation sum rule of PDFs, and, in the forward limit, the sum rule for $H_5$ in Eq. (\ref{eq:sum-rules-GPD-FF-45}) recovers the parton model sum rule $\int b_1(x) dx =0$ \cite{Close:1990zw}.

\section{Summary}
\label{sec:Discussion}
In this paper, we formulate the EMT form factors for a spin-1 hadrons. The energy density, spin distribution and stress tensor are given. 
The pressure and shear forces functions are found in terms of multipole expansion as the spin-1 particle is not spherically symmetric.
The sum rules between the GFFs and GPDs are derived.\\
%and, especially, the quark spin contribution is separated into the usual quark helicity (spin part) plus the gauge-invariant orbital angular momentum in the similar way as the spin 1/2 case \cite{Ji:1996ek}, which can help us to learn more about the spin structure of spin 1 hadrons. \\

\noindent
{\bf Note added}\\

During finishing the present manuscript we became aware of recent Ref.~\cite{Cosyn:2019aio} where the EMT form factors for spin-1 particles were also considered.

\begin{acknowledgments}
We are grateful to Cedric Lorce, Kirill Semenov-Tian-Shansky, and Peter Schweitzer
 for illuminating 
discussions and inspiration.
This work was supported  by CRC110 (DFG) and the State Scholarship Fund No. 201804910428 (CSC) and the DAAD Research Grants No. 57381332.
\end{acknowledgments}
\appendix

\section{Breit frame formulae}
\label{sec:apdix-gffs}

The matrix elements of the spin 1 quadrupole operator in Eq. \ref{eq:quadrupole}, 
\eq \label{eq:quadrupole-2}
\hat Q^{ik}_{\sigma'\sigma}&=& \frac 12 \left( \hat S^{\ i} \hat S^{\ k}+\hat S^{\ k} \hat S^{\ i} -\frac43 \delta^{ik} \right)_{\sigma'\sigma} \nonumber \\
&=& \frac{1}{3} \delta_{ij} \, \delta_{\sigma'\sigma} - \frac{1}{2} \left( \hat{\epsilon}_{\sigma'\, j}^* \hat{\epsilon}_{\sigma \, i} + \hat{\epsilon}_{\sigma'\, i}^* \hat{\epsilon}_{\sigma \, j}  \right) 
\en
where $\sigma=\{+,-,0\}$, and 
\eq
{\hat S}^i {\hat S}^j - {\hat S}^j {\hat S}^i = i \ep^{ijk} {\hat S}^{\,k} .
\en

In Breit frame, the initial(final) momentum $p^\mu$($p^{\prime\mu}$) has the relation $P^\mu=(p^\mu+p^{\prime\mu})/2=(E,0,0,0)$ and $\Delta^\mu=p^{\prime\mu}-p^\mu=(0,{\vec \Delta})$. So ${\vec p}=-{\vec p}^{\,{ }\prime}=-{\vec \Delta}/2$ and $p^0=p^{\prime\, 0}=E=\sqrt{m^2-t/4}$ with $t=\Delta^2$. So initial and final polarizition vectors are 
\eq \label{eq:polarization_vector_2}
\ep^\mu ({p},\sigma) 
&=& \left( - \frac{\vec{\Delta}\cdot\hat\ep_\sigma}{2m}, \hat\ep_\sigma + \frac{\vec{\Delta}\cdot\hat\ep_\sigma}{4m(m+{E})}\vec{\Delta}\right),  \\
\ep^\mu ({p'},\sigma') 
&=& \left( \frac{\vec{\Delta}\cdot\hat\ep_{\sigma'}}{2m}, \hat\ep_{\sigma'} + \frac{\vec{\Delta}\cdot\hat\ep_{\sigma'}}{4m(m+{E})}\vec{\Delta}\right),  
\en
So one can get the following useful relations (here ${\hat Q}_{kl}={\hat Q}^{kl}_{\sigma'\sigma}$ is the matrix element and $\ep^{\prime* \mu}=\ep_{{\sigma'}}^{*\mu}=\ep^{*\mu} ({p'},\sigma') $, $\ep^\mu=\ep_{\sigma}^\mu=\ep^\mu ({p},\sigma)$, and note $t=\Delta^2$),
\sub{
\label{eq:epep}
\eq 
%&& {\text{Eq. (\ref{eq:epB_D_epA_D_sij}): }} \quad 
(\hat\ep_{\sigma'}^* \cdot \vec \Delta)(\hat\ep_{{\sigma'}} \cdot \vec \Delta) 
&=& -{t\over3}  \delta_{\sigma'\sigma} - {\hat Q}_{kl} { \Delta}_k { \Delta}_l  \ ,   \\ [0.7em] 
%
%&&{\text{Eq. (\ref{eq:epBdotepA_sij}): }} \quad 
\ep^*_{\sigma'} \cdot \ep_{\sigma}
&=& \left( { t \over 6m^2} -1 \right)   \delta_{\sigma'\sigma}  + { 1 \over 2m^2} {\hat Q}_{kl}  { \Delta}_k { \Delta}_l   \ ,   \\[0.7em]
%
%&& {\text{Eq. (\ref{eq:epB0epA0} ): }} \quad 
\ep_{{\sigma'},0}^*\ep_{{\sigma'},0} 
&=&  { t \over 12m^2}  \delta_{\sigma'\sigma} 
+ { 1 \over 4m^2} {\hat Q}_{kl}  { \Delta}_k { \Delta}_l 
\,  \\  [0.7em]
 % 
%&&{\text{Eq. (\ref{eq:vec_epA_D} ): }} \quad 
\ep_{\sigma} \cdot \Delta 
&=& -{E\over m} \, \hat\ep_{\sigma} \cdot \vec \Delta \ ,   \\ [0.7em]
%
%&&{\text{Eq. (\ref{eq:vec_epB_D} ): }} \quad 
\ep_{\sigma'}^* \cdot \Delta &=& -{E\over m} \, \hat\ep_{\sigma'}^* \cdot \vec \Delta  \ ,  \\[0.7em]
%
%&&{\text{Eq. (\ref{eq:vec_epB_D_epA_D} ): }} \quad 
(\ep_{\sigma'}^* \cdot \Delta)(\ep_\sigma \cdot \Delta) 
&=& - {E^2\over m^2}  \left( {t\over3}   \delta_{\sigma'\sigma}  + {\hat Q}_{kl}  { \Delta}^k { \Delta}^l  \right) \ ,  \\ [0.7em]
%
%&&{\text{Eq. (\ref{eq:vec_epB0_epA_D} ): }} \quad 
\ep_{{\sigma'},0}^* \ep_\sigma \cdot \Delta  
&=& 
{E \,  \over 2m^2} \left( {t\over3}   \delta_{\sigma'\sigma} + {\hat Q}_{kl} { \Delta}^k { \Delta}^l  \right) \ , \\ [0.7em]
%
%&&{\text{Eq. (\ref{eq:vec_epA0_epB_D} ): }} \quad 
\ep_{\sigma,0} \ep_{\sigma'}^* \cdot \Delta 
&=& -\ep_{{\sigma'},0}^* \ep_\sigma \cdot \Delta \ ,  \\ [0.7em]
%
%&&{\text{Eq. (\ref{eq:vec_epBj_epA_D} ): }} \quad 
\ep_{{\sigma'},j}^* \ep_\sigma\cdot \Delta - \ep_{\sigma,j}\ep_{\sigma'}^*\cdot \Delta
&=& {iE\over m} \Delta_k \ep^{kjl} {\hat S}^l_{{\sigma'}{\sigma}}  \\[0.7em]
%
%&&{\text{Eq. (\ref{eq:vec_epA0_epBj} ): }} \quad 
\ep_{\sigma,0}\ep_{{\sigma'},j}^*+\ep_{{\sigma'},0}^*\ep_{\sigma,j} 
&=& {  i \over 2m} \Delta_k  \ep^{kjl} {\hat S}^l_{{\sigma'}{\sigma}} \\[0.7em]
%
%&& {\text{Eq. (\ref{eq:vec_epAi_epBj} ): }} \quad 
\ep_{{\sigma},i}\ep_{{\sigma'},j}^*+\ep_{{\sigma'},i}^*\ep_{{\sigma'}j} 
&=& \left( {2\over3} \delta_{ij} +{1 \over 6m^2} \Delta_i \Delta_j \right) \delta_{\sigma'\sigma}
- 2 {\hat Q}_{ij}  
-{1\over2m(m+E)} \left( { \Delta}_i { \Delta}_k {\hat Q}_{kj}  
+ { \Delta}_j { \Delta}_k \hat Q_{ki}  \right) 
\nonumber \\ 
&&
- {1\over8m^2(m+E)^2} { \Delta}_i { \Delta}_j 
 { \Delta}_k { \Delta}_l {\hat Q}_{kl}
\en  }

%\bibliography{ref}

\end{document}